\documentclass{article}
\usepackage{url}
\usepackage[margin=1in]{geometry} 
\usepackage{amsmath,amsthm,amssymb,amsfonts}
\usepackage{graphicx}
\usepackage{booktabs}
\usepackage{mathtools}
\usepackage{lineno}
\usepackage{float}
\usepackage{caption,subcaption,setspace}
\usepackage[bottom]{footmisc}

\title{Heuristic assessment of the economic effects of pandemic control}

\author{
Xiang Niu\thanks{Network Science and Technology Center, Rensselaer Polytechnic Institute (RPI), Troy, NY 12180, USA} \thanks{Department of Computer Science, Rensselaer Polytechnic Institute (RPI), Troy, NY 12180, USA} 
\and 
Christopher Brissette\footnotemark[2] 
\and
Chunheng Jiang\footnotemark[1] \footnotemark[2] 
\and
Jianxi Gao\footnotemark[1] \footnotemark[2] 
\and
Gyorgy Korniss\footnotemark[1] \thanks{Department of Physics, Rensselaer Polytechnic Institute (RPI), Troy, NY 12180, USA} 
\and
Boleslaw K. Szymanski\footnotemark[1] \footnotemark[2] \footnotemark[3] \thanks{Corresponding author: boleslaw.szymanski@gmail.com} }

\begin{document}

\maketitle
\doublespacing
\section{Abstract}
Data-driven risk networks describe many complex system dynamics arising in fields such as epidemiology and ecology. They lack explicit dynamics and have multiple sources of cost, both of which are beyond the current scope of traditional control theory. We construct the global risk network by combining the consensus of experts from the World Economic Forum with risk activation data to define its topology and interactions. Many of these risks, including extreme weather, pose significant economic costs when active. We introduce a method for converting network interaction data into continuous dynamics to which we apply optimal control. We contribute the first method for constructing and controlling risk network dynamics based on empirically collected data. We identify seven risks commonly used by governments to control COVID-19 spread and show that many alternative driver risk sets exist with potentially lower cost of control.


\section{Introduction}
\noindent Network dynamics define a plethora of real world systems and describe the evolution of various spreading processes ranging from epidemiology to economic shocks. In recent years ideas from control theory have been adapted to network science with some success in understanding structural linear control of static networks~\cite{liu2011controllability,yan2015spectrum,yan2017network,nepusz2012controlling,ruths2014control}, temporal networks~\cite{li2017fundamental,posfai2014structural,paranjape2017motifs}, and multilayer networks~\cite{posfai2016controllability}. Current work also focuses on networks with nonlinear dynamics~\cite{zanudo2017structure,cornelius2013realistic}. Comparatively little work studies the evolution and control of risk networks where active risks incur additional cost to the controller beyond the cost of the input signal~\cite{helbing2013globally}. A differentiating factor of risk networks in the context of control is that the activity of a single risk has the capability of activating others in a potentially catastrophic cascade that makes component-oriented analysis of individual risks inadequate for understanding their consequences. To fully understand the prevailing risks around us, we must understand the ways in which those risks interact and the ways those interactions evolve yielding risk dynamics. Despite a large body of literature on risk management and mitigation~\cite{tait1992proactive,erisman2015global,who2016after}, the fundamental questions of networked risk control are rarely discussed~\cite{helbing2013globally,szymanski2015failure} as they are beyond the state of the art in both control engineering and network science. \\
\indent Two problems facing the study of networked risk control are the need for explicit dynamics and optimizing over multiple cost types. The former arises from the fact most real-world dynamics are informally defined by data as opposed to the formal differential equations required for control theory. We address this problem by defining a stochastic process that transitions each risk among a finite set of states such as active, non-active, or recovering. Collectively these stochastic processes define our dynamics. In Figure 1 we show how activity in a subnetwork of the larger global risk network changes over time when nodes take a binary state and are allowed to be either active or inactive. 

\begin{figure}[h!]
    \centering
    \begin{subfigure}[b]{\textwidth}
        \includegraphics[width=\textwidth]{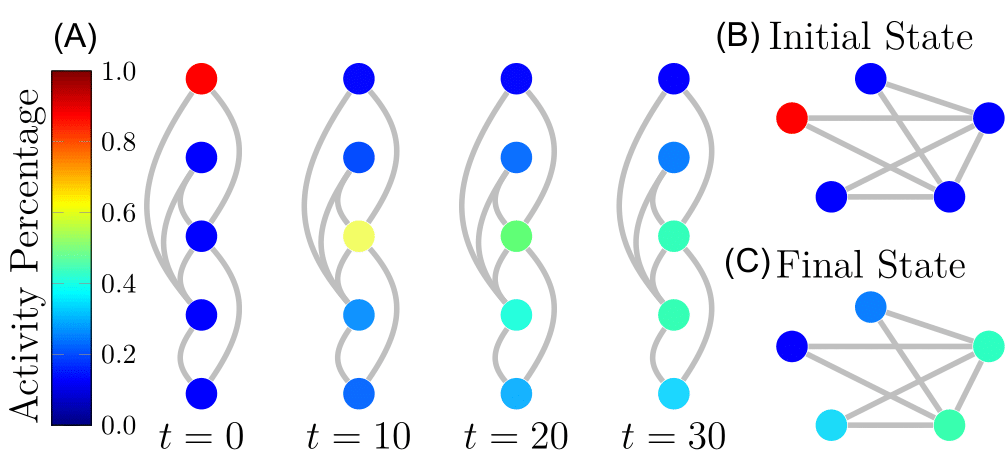}
        \label{fig:dynamicrisk}
    \end{subfigure}
    \caption{An example of continuous risk dynamics on a subnetwork from the World Economic Forum's Global Risk Network consisting of inflation, failure of climate change mitigation, interstate conflict, large scale migration, and cyberattacks. No control is applied and we can see that all nodes are inactive at the beginning of the simulation except for inflation. The activity from inflation can be seen to disperse over the network until all nodes are at a low level of activity at the end of the simulation.}
    \end{figure}   
    
We adapt these states to represent a continuous dynamical analogue of the discrete dynamics in order to apply optimal control theory~\cite{todorov2002optimal,sideris2005efficient,li2012controllability}. The equations we consider controlling take the following general form:
\begin{equation}
    \vec{x}(k+1) = F[\vec{x}(k)] + G[\vec{x}(k),E] + B\vec{u}(k)
\end{equation}

\noindent where $x_i(k)$ is the expected value of risk $i$ at time step $k$, $F[\vec{x}(k)]$ and $G[\vec{x}(k),E]$ are nonlinear functions depending on the endogenous and exogenous probabilities of activation for risks respectively, $E$ is the adjacency matrix defining interactions between pairs of risks, and $B\vec{u}(k)$ defines our input control signals. Using the above continuous equation, we apply an altered version of the linear-quadratic regulator to account for multiple cost types and ensure optimal control. We apply these methods to the annually published global risk network from the World Economic Forum in order to obtain the dynamic global risk network and control it. The 2017 WEF global risk network can be seen in Figure 2. 

\captionsetup[subfigure]{labelformat=empty}
    \begin{figure}[!ht]
    \centering
        \begin{subfigure}[b]{0.8\textwidth}
            \caption{(A)}
            \centering
            \includegraphics[width=0.9\textwidth]{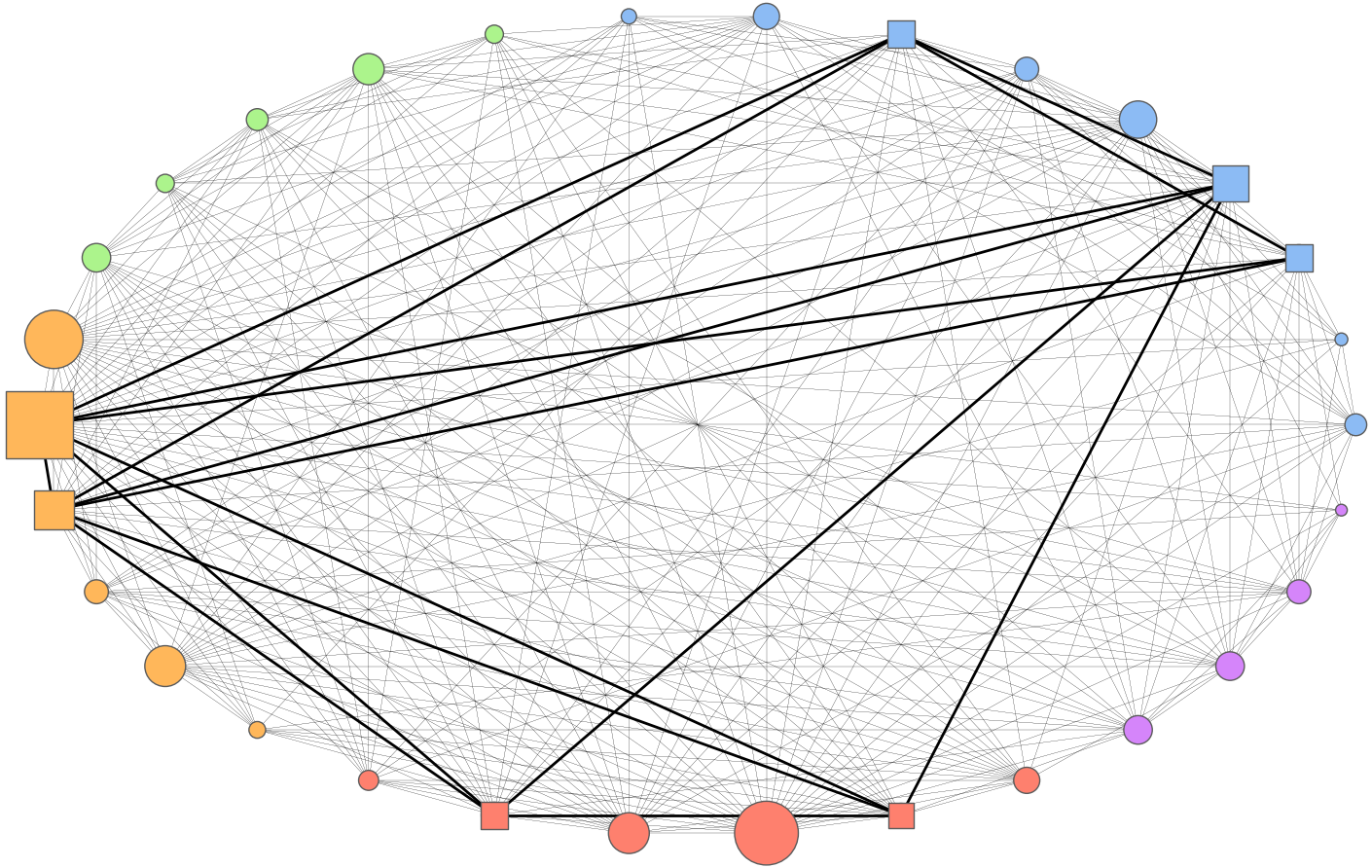}
            \includegraphics[width=\textwidth]{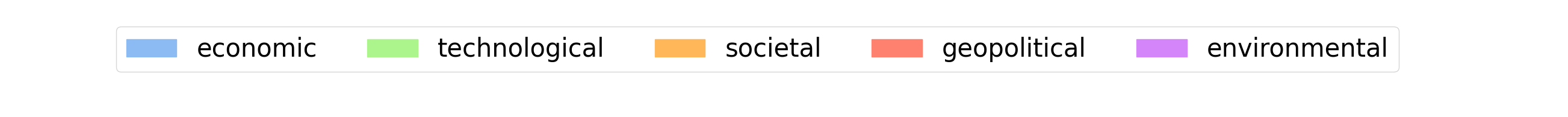}
        \end{subfigure}
        \begin{subfigure}[b]{0.8\textwidth}
        \begin{subfigure}[b]{0.53\textwidth}
            \caption{(B)}
            \includegraphics[width=\textwidth]{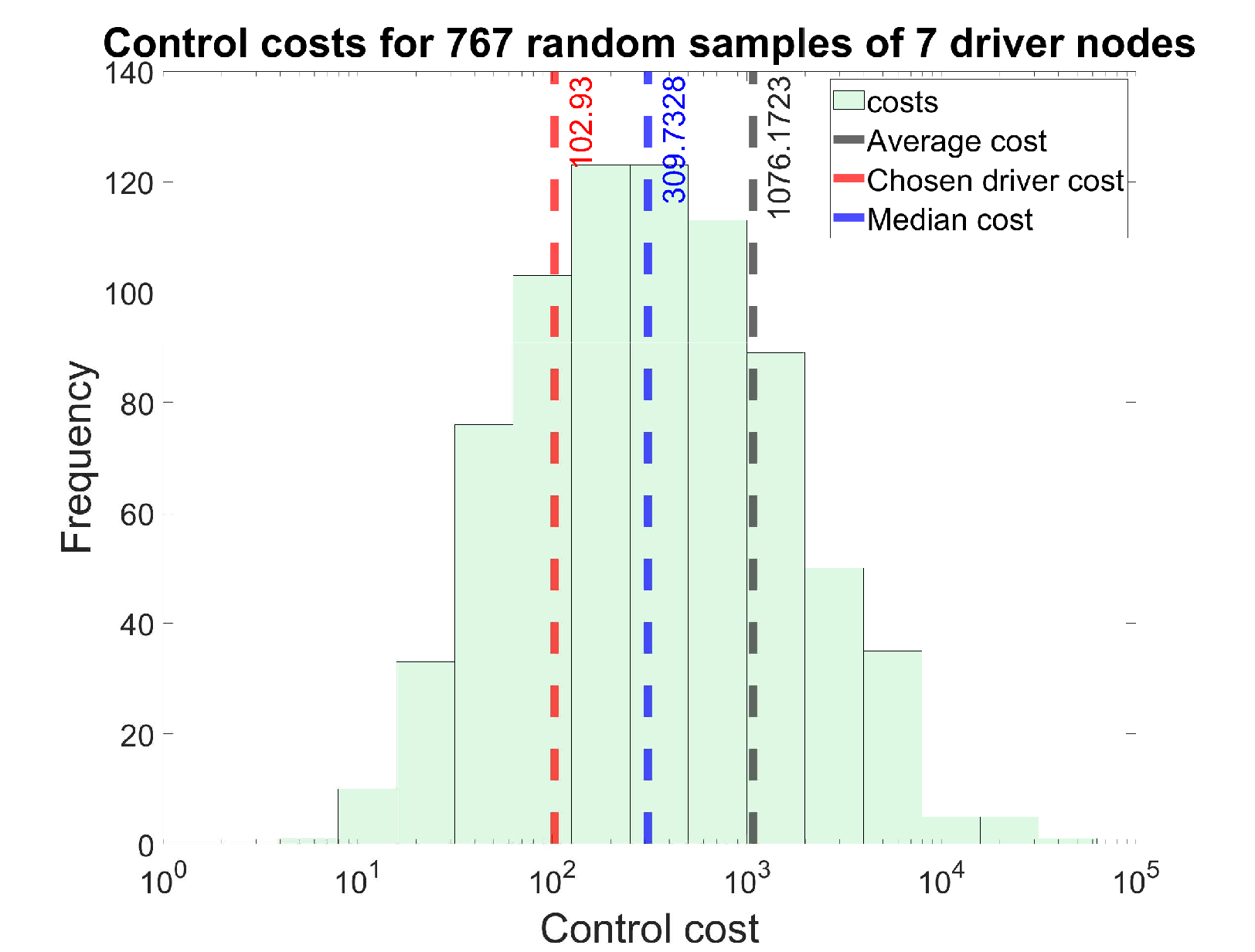}
        \end{subfigure}
        \begin{subfigure}[b]{0.5\textwidth}
            \caption{(C)}
            \includegraphics[width=\textwidth]{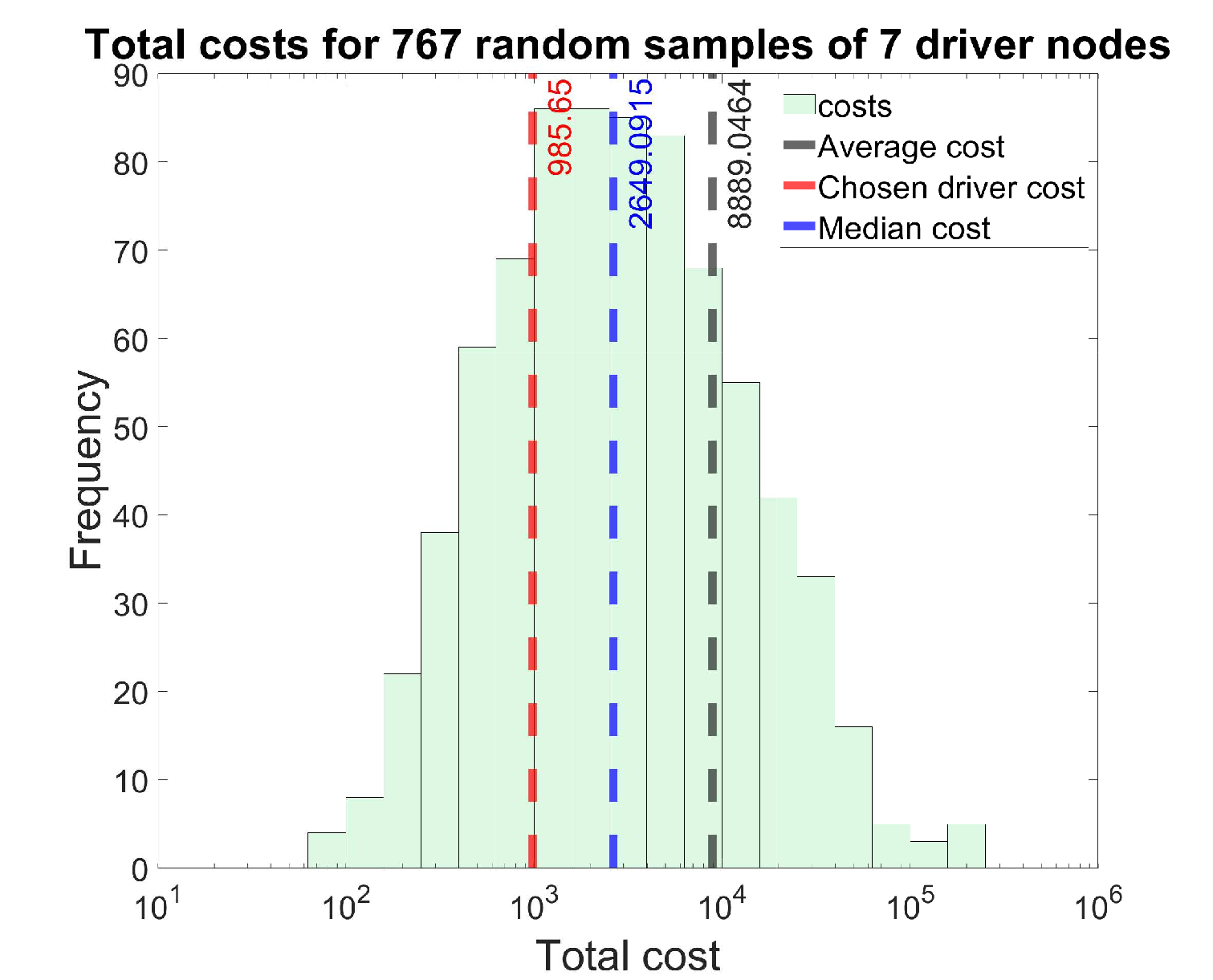}
        \end{subfigure}
        \label{fig:drivers}
        \end{subfigure}

    \caption{Incurred cost from optimal control for random samples of 7 driver nodes on the 2017 dynamic risk network with the "rapid and massive spread of infectious diseases" risk held constant at one. The network itself is dense and near regular with mean vertex degree of 18.27 and standard deviation of 4.60. We compare the 7 node costs with that of the nodal set consisting of deflation, failure of major financial institutions, unemployment, failure of national governance, failure of global governance, failure of urban planning, and profound social instability. These risks relate closely with observed government action in response to COVID-19. In the above network we have highlighted the nodes within the 2017 global risk network these drivers consist of by making them square. This driver set performs better than average by a reasonable margin, but we also see that many other driver sets performed better.}
    \end{figure}

\indent Globalization has provided extensive quality of life improvements for billions of people worldwide, including increased life expectancy, poverty reduction, and far reaching medical advances. Also due to globalization, risks are now more connected through the avenues of technology, business, and individuals than ever. As both the global economy and the technology defining our interactions become more connected, they are also becoming more vulnerable~\cite{wef2020grr,helbing2013globally}. The annually reported global risk network from the World Economic Forum~\cite{helbing2013globally} currently represents the best understanding of the ways systemic risks are connected to each other. This network distills the immeasurably complex and constantly evolving network of global economic factors based on a consensus of expert opinions, and instead includes various transitions to catastrophic states such as unmanageable inflation~\cite{pindiriri2012monetary}, climate change~\cite{lobell2008prioritizing}, interstate conflict~\cite{lee2003explaining}, involuntary migration~\cite{mcgregor1994climate}, and large-scale cyber attacks~\cite{cashell2004economic}, a subnetwork of these risks can be seen in Figure 1.
    
While active, corresponding risks induce tremendous damages to resources, economies, and most importantly human lives. As examples consider the 2014 Ebola virus outbreak in Africa~\cite{piot2014ebola,who2018ebola,dellicour2018phylodynamic} and the
2008 financial meltdown triggered by a crisis in the subprime mortgage market~\cite{morris2008financial,reinhart20082007,ivashina2010bank,battiston2016complexity}. Neither the viral outbreak nor the financial meltdown were confined to where they erupted nor limited to initially triggered risks, but continued to grow and affect other risks, often causing enormous losses. An even more contemporary example of networked risk dynamics comes from the 2019 breakout of COVID-19. In response to the epidemic, nations around the world enacted a plethora of control measures to mitigate the virus's effects. While policies deviated between countries, control measures generally included the temporary closure of businesses, stimulus packages, and travel limitations. It is still unknown what the long term fallout of COVID-19 will be, and understanding the risks associated with the disease is going to be important in addressing long term effects. According to the Global Bank COVID-19 could lead to widespread school closings, increased dropout rates, and decreased development in human capital in Europe and Central Asia~\cite{gb2020covid}. In South America, the Caribbean, and South Africa the Global Bank suspects social unrest may arise due to food shortages related to COVID-19~\cite{gb2020covid}. In addition to this analysis, the World Economic Forum also released a report on risks associated to COVID-19 based on the risks outlined in their yearly Global Risk Report~\cite{wef2020covid}. They listed prolonged recession of the global economy, high levels of structural unemployment, tighter restriction on cross-border movement of people and goods, and economic collapse of emerging markets and developing economies as the most prevalent risks.\\

\section{Results}
\subsection{The dynamic global risk network}
\noindent The simplified abstraction of the dynamic global risk network~\cite{niu2018evolution} indicates connections between major risks that are prone to cascading failures. These risks are active or inactive and evolve as a Markov process depending on the network's current state. This poses a significant challenge to control theory, since applying theory from control analysis requires continuously defined dynamics for the risk variables. Therefore, we introduce a continuous dynamical model describing how risk activation propagates through the underlying network.  We model the network activity dynamics with an adaptation of a stochastic process that is based on an alternating renewal process~\cite{cox1965theory,szymanski2015failure,yao2012uncertain,majdendzic2014spontaneous2014} and is continuous. Instead of each risk transitioning to a discrete state indicating activity, we have the state of each node represent the expected value of risk $i$ being active at time step $k+1$, $x_i(k+1)$ using the following model~\cite{szymanski2015failure,lin2017limits, niu2017evolution,moussawi2018carp,niu2018evolution}.

\begin{equation}
    \vec{x}(k+1) = F[\vec{x}(k),\vec{p}_{int},\vec{p}_{con}] + G[\vec{x}(k),\vec{p}_{ext},E] + B\vec{u}(k)
\end{equation}

\indent where $B$ is the diagonal binary matrix representing the family of nodes being driven by the control signal vector $\vec{u}(k)$. Both $F[\vec{x}(k),\vec{p}_{int},\vec{p}_{con}]$ and $G[\vec{x}(k),\vec{p}_{ext},E]$ are nonlinear dynamical functions depending on the state vector $\vec{x}(k)$. Here the probabilities $\vec{p}_{int}$, $\vec{p}_{ext}$, and $\vec{p}_{con}$ represent the internal and external probabilities of activation, and the probability nodes retain their state, respectively. Each of these probabilities is calculated according to a maximum likelihood estimation as discussed in previous work~\cite{lin2017limits}. The function $G$ is additionally dependent on the underlying network defined by the possibly weighted adjacency matrix $E$.

\indent Since the individual states $x_i(k)$ at the given time $k$ vary in meaning for different applications, so too do the signal costs, $u_i(k)$. For example in the case of a Lotka Volterra network an individual control signal $u_i(k)$ may represent additional population added or removed at time $k$, where the state $x_i(k)$ represents the population of a given species at that time. Alternatively in a mechanical system $x_i(k)$ may represent quantities such as velocity, or force of moving components, while $u_i(k)$ represents energy applied to the system. In the case of the Lotka Volterra network the nonlinear dynamics $F[\vec{x}(k),\vec{p}_{int},\vec{p}_{con}]$ and $G[\vec{x}(k),\vec{p}_{ext},A]$ may instead represent the growth rate of a node's population based on its current population and its neighbor's populations respectively. Similarly, for a mechanical system the equations may represent the resistance of a component and the force being applied to it by other components respectively. 

\indent We note that while this model is convenient and provides accurate risk activity estimates, real risk networks are constantly evolving. Risks can be added to the network as they arise and they may also be removed due to factors such as government and industries intervention. Consequently the underlying network changes and as such, the probabilities associated with each risk change to accommodate fluctuations in network interactions. This implies the risk network and its underlying dynamics should be regularly revisited and redefined. 

\subsection{Controlling risk networks}
\noindent When applying control theory to nonlinear problems, we require a linearization of the underlying nonlinear dynamics. Assume that when uncontrolled, the system in equation (2) approaches a natural steady state $\vec{x}_s$. Also assume $f_x = F + G$ and $A = \frac{\partial{f_x}}{\partial{\vec{x}}}\vert_{\vec{x}=\vec{x}_s,\vec{u}=\vec{u}_s}$ is the adjacency matrix defining the underlying probability of activation between links. Then we linearize (1) as follows.

\begin{equation}
    \Delta\vec{x}(k+1) \approx A\Delta\vec{x}(k) + B\Delta\vec{u}(k)
\end{equation}

Here $\Delta\vec{x}(k) = \vec{x}(k)-\vec{x}_s$ and $\Delta\vec{u}(k) = \vec{u}(k)-\vec{u}_s$.

\indent If the linearized system (3) is locally controllable along a specific trajectory then the associated non-linear system is controllable along the same trajectory. Therefore the linearized system is sufficient for several parts of control analysis including determination of driver nodes and determination of instanaeous optimal control. In traditional control, the control energy after $\tau$ time steps is expressed as a sum over our control signal at each time step given by $J_{\epsilon}=\overset{\tau-1}{\underset{k=0}{\sum}}||\vec{u}(k)||^2$.
$J_{\epsilon}$ depends only on the control signal, whereas we require a cost function additionally depending on risk activity. For this purpose, we alter the cost function of the linear quadratic regulator to obtain the following control cost which includes the cost of active risks.

\begin{equation}
    J_t(Q_f,Q,R) = \vec{x}^T(\tau)Q_f\vec{x}(\tau) + \overset{\tau - 1}{\underset{k=k_0}{\sum}}[\vec{x}^T(k)Q\vec{x}(k)+\vec{u}^T(k)R\vec{u}(k)]
\end{equation}

Here $Q_f$ denotes the cost matrix at the final state, $Q$ is the cost matrix for intermediate states, and $R$ is the cost matrix for our control signals. Examining this equation we see that the right most term is a sum over the costs associated with the state of each risk and the control cost being applied to each risk. For our tests we assume all of these matrices to be the identity for simplicity. As such, we have expressed the cost of controlling our risk network in a form where we may apply an optimal control strategy to reduce the overall incurred cost. 

\indent In the case of networked risks, system dynamics have a natural steady state we will refer to as $\vec{x}_s$. This is the state that the system will approach in the absence of a control signal. Because of this, when trying to drive the entire network towards an alternate state $\vec{x}_* \neq \vec{x}_s$ it is necessary to continuously apply a control signal. Therefore, we introduce a distinction between reactive and proactive control. It is important to note that this is common in risk reduction literature~\cite{tait1992proactive}, however it lacks theoretical or heuristic tools for assessment. To motivate the need for heuristic cost estimates we examine the case of controlling the activity of COVID-19 within the 2017 dynamic risk network. In order to better understand the hidden costs associated with control measures related to COVID-19 we construct a set of seven control nodes within the 2017 risk network associated with observed government intervention and use optimal energy control to attempt driving the dynamic global risk network towards inactivity. The nodal set consists of deflation, failure of major financial institutions, unemployment, failure of national governance, failure of global governance, failure of urban planning, and profound social instability. The cost of control measures on this driver set is compared with the cost of controlling the network using seven randomly sampled control nodes. The results of this simulation are demonstrated in Figure 2. We find that while the seven nodes chosen incur relatively low cost among the sample, there are many more efficient sets of control nodes one could pick. This suggests that the policy decisions made in response to COVID-19 were reasonable ones, but not optimal. It should be noted that the costs associated with each risk are taken to be unit while in reality they would likely vary and require expert opinion to deduce. Because the costs are all unit, this method is giving us a qualitative understanding of how topologically important our chosen control set is within the dynamic risk network. This highlights the combinatorial nature of finding optimal risk driver sets and the need for heuristic assessment tools in choosing them. Unfortunately such heuristics are not widespread and the current understanding of risk assessment is often limited to individual, or narrow groups of risks~\cite{winsemius2016global,o2017ipcc,hoekstra2014water,wada2014wedge,manning2010misrepresentation,vorosmarty2015scale,pekel2016high,mekonnen2016four,maynard2015nano,liang2012sexual}. 

\subsection{Assessing control methods}
\noindent We suggest two heuristics to help inform driver node choice when applying reactive and proactive control phases respectively on risk networks. Call $N_D$ the total number of driver nodes, $N_{D_a}$ the number of driver nodes that are initially active, and $N_{D_p}$ the number of driver nodes that are among the most active at the system's natural steady state $\vec{x}_s$. In Figure 3, we see how $N_{D_a}$ affects both the control cost and total cost in the reactive control phase. For the reactive phase we see a strong negative relation between the control cost and the total cost. 

\begin{figure}[!ht]
    \centering
        \begin{subfigure}[b]{\textwidth}
        \begin{subfigure}[b]{0.5\textwidth}
            \caption{(A)}
            \includegraphics[width=\textwidth]{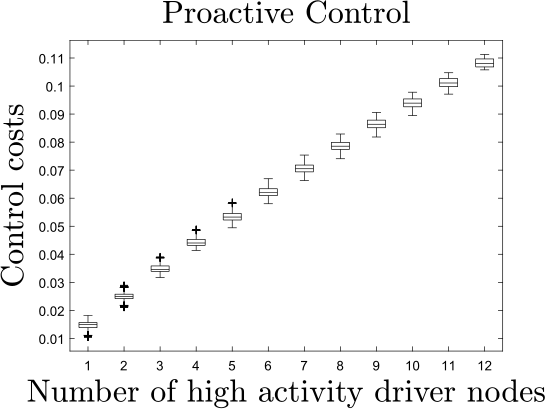}
        \end{subfigure}
        \begin{subfigure}[b]{0.5\textwidth}
            \caption{(B)}
            \includegraphics[width=\textwidth]{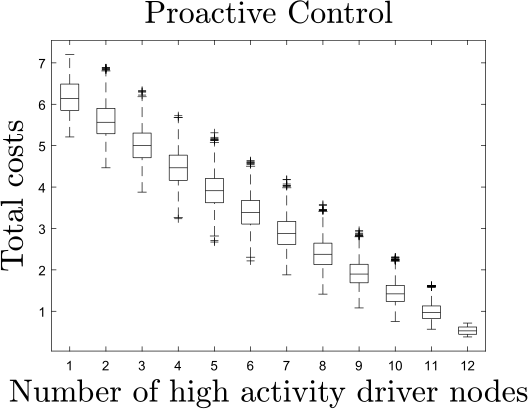}
        \end{subfigure}
        \label{fig:proactive}
        \end{subfigure}
    \centering
        \begin{subfigure}[b]{\textwidth}
        \begin{subfigure}[b]{0.5\textwidth}
            \caption{(C)}
            \includegraphics[width=\textwidth]{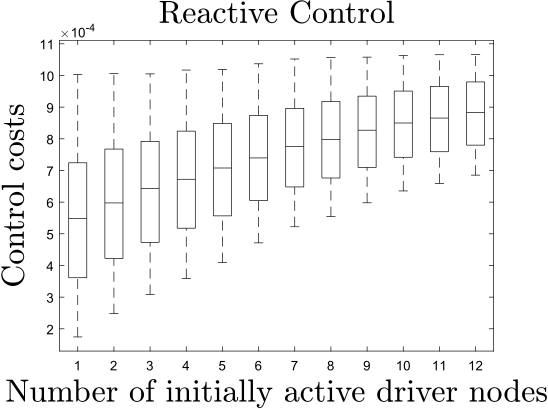}
        \end{subfigure}
        \begin{subfigure}[b]{0.5\textwidth}
            \caption{(D)}
            \includegraphics[width=\textwidth]{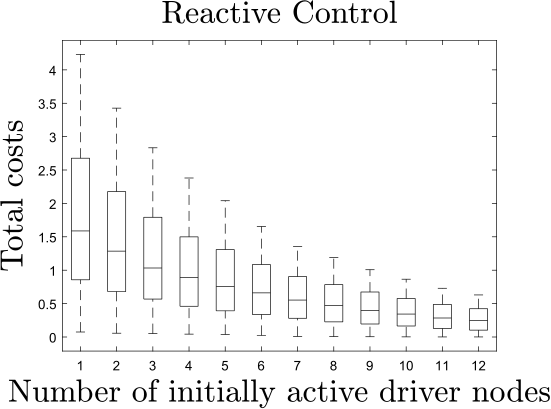}
        \end{subfigure}
        \label{fig:reactive}
        \end{subfigure}
    \caption{We show the relationships between the number of "high impact" nodes in our control set $N_{D_a}$ and $N_{D_p}$ and the effects on costs incurred during the reactive and proactive control phases respectively. There were 12 total control nodes in all tests and for each associated $N_{D_a}$ and $N_{D_p}$ 100 driver node sets were sampled. In the proactive phase the network was controlled for 50 time steps, and in the reactive phase the network was controlled for 500 time steps. We can see that control costs went up with an increase in $N_{D_a}$ or $N_{D_p}$ in both the reactive and proactive control phases respectively. Alternatively we see the opposite trend in total costs. Total costs appear to decline far more with $N_{D_p}$ in the proactive phase than they do with $N_{D_a}$ in the reactive phase.}
    \end{figure}

\indent Once the system has been driven to inactivity, the dynamics will continue to evolve towards the natural steady state of the system $\vec{x}_s$. To combat this, we use proactive control in which we apply a control signal to our driver nodes based on their activation probabilities at the natural steady state $\vec{x}_s$. The cost of this control phase depends on $N_D$ just as the reactive phase did. We also find that the number of driver nodes with high activation probabilities at the natural steady state $\vec{x}_s$, $N_{D_p}$ has a drastic effect on the control cost.  In Figure 3, we plot the control costs and total costs against each other for collections of driver nodes with different $N_{D_p}$ in the proactive control phase. We can see that the trends are broadly similar to those seen in the reactive phase with respect to $N_{D_a}$. We see that a high number of $N_{D_p}$ among our driver nodes generally reduces our total cost and increases our control cost. However, the control cost and total cost here have far smaller quartiles than in the reactive phase seen in Figure 3. 

\indent It should be noted that applying this control to real-world networks is not a trivial problem in and of itself. In the case of the Global Risk Network, the control signal must be designed by experts and may take the form of strategies such as enacting legal policy, investing, or quarantining infections. In practice, it may require iterations of design to force risks towards inactivity, and in that time, the underlying network and probabilities defining its dynamics may change. Despite this, knowing which set of risks forms an optimal set of drivers is valuable in its own right and can minimize control costs.\\

\section{Discussion}
\noindent Networked risks provide a theoretical foundation for defining complex interactions between factors that are consistently prone to cascading failures. To avoid the damages of inevitable steady states that arise from the interactions of these networks, we require an optimal method to them. The dynamics resulting from these networks are difficult to analytically define for use in control theory since they must be constructed from probabilities and extensive data collection. Our method presents a pipeline for constructing dynamic risk networks from extensive data and how to control them. This requires using a massive amount of collected data and applying maximum likelihood estimation in order to predict transition probabilities. Using these transition probabilities, we can construct continuous dynamics from the alternating renewal process~\cite{cox1965theory} that defines the network's underlying discrete dynamics. In these new continuous dynamics, the state of each node represents its probability of activity over the current time step as opposed to the initially defined discrete dynamics. These continuous dynamics allow for the application of control methods for driving the system into inactivity. We adapt the linear quadratic regulator to account for risk networks in which control cost and risk activity cost are simultaneously considered. We also show that by altering the control strategy before and after driving the network to inactivity we can drastically reduce the total cost of controlling the system. 

\indent The tools proposed in this paper are very general and widely applicable. It should be noted that a trade-off between proactive and reactive control arises not only in risk networks but in any system in which the desired final state of control is not stable. Most of the control designs for such systems make a salient assumption that the cost of the system being out of the desired final state is negligible. Certainly, there are other systems than risk networks, in which this assumption is not true. Our approach to control such systems in two phases, reactive and proactive, can be applied to such cases. Hence the usefulness of our approach reaches beyond risk networks.

\section{Limitations of the Study}
\noindent We first note that this model requires consistent reevaluation from experts. Both the connectivity and the weighting of the links in the dynamic risk network are subject to change as experts reevaluate and add new risks. Additionally the cost matrices used in our model are subject to change with expert evaluation as well. Furthermore, applying the linear quadratic regular as an explicit control method to the dynamic global risk network is difficult in practice. The control signal being added to nodes in our control set varies over many professional domains and would realistically require the fine tuning of many distinct policy decisions. For this reason, in many real world applications we suggest that this method be used as a heuristic for evaluating comparative costs between driver node sets as opposed to an explicit method for generating real-world control strategies.

\section{Methods}
\subsection{Experimental Design}
This study aims to develop a heuristic understanding of which nodes are optimal for controlling risk networks. We specifically study this on the dynamic global risk network derived from the World Economic Forum’s global risk network. Our study takes random samples of driver nodes and simulates the networked dynamics while assess the control costs of forcing the network to inactivity. 
\subsection{CARP Model}
In the Cascading Alternating Renewal Processes (CARP) model, a node $i$ has binary state $x_i(k)$ at each step $k$, either active (1), or inactive (0).
In the global risk networks, an active node state represents a materialized risk.
The model is Markovian, so at each time step, a node
computes for each node its state for the next time step based on states of nodes at the current time step.
There are three transitions: 
\begin{itemize}
	\item
	Internal activation: an inactive node $i$ is activated internally with probability $p_i^{int}$.
	\item
	External activation: an inactive node $i$ is activated externally by an active node $j$ with probability $p_{ji}^{ext}=p_{i}^{ext}$.
	\item
	Internal recovery: an active node $i$ is inactivated internally with probability $p_{i}^{rec}=1-p_{i}^{con}$.
\end{itemize}
The dynamics of CARP model is defined by the following transitions:
\begin{align*}
\small
\label{eq_risk_nonlinear_updates}
\vec{x}(k+1) &= f_{x}^{\rm CARP}[\vec{x}(k)] \\
&= \vec{x}(k+1)=\{\vec{p}^{int}+[E^{\intercal}\vec{x}(k)\circ\vec{p}^{ext}]\}\circ[1-\vec{x}(k)]+\vec{p}^{con}\circ\vec{x}(k)
\end{align*}
where $E_{ij}\in\{0,1\}$ and $E_{ij}$ is 1 if and only if edge $(i, j)$ exists. It should be noted that both risks themselves and their probabilities of transitions between states  constantly change; new risks arise and are added to the network, while existing active risks either continue to be a threat and remain in the network but with modified parameters, or, thanks to the response of threatened governments and industry, decline in importance and are removed. This evolution causes continuous changes in the global risks and their probabilities, and leads to the need for annual revisions of the list of risks present in the network and their parameters by the expert. However, if left unabated, the global risk network would approach the steady state, in which some risk will be active much more frequently making their threats much more pronounced than in the initial state. 
\subsection{Statistical analysis}
The The histograms in Figure 2 (B) and (C) come from simulations run on a random sample of 767 7-node driver sets driving the dynamic global risk network. In these experiments the node associated with the spread of infectious disease was held constant and the control method being used is optimal energy control. \\
\indent The subfigures in Figure 3 each come from a sample of 1200 driver node sets. Each driver set consists of 12 nodes. For subfigures (A) and (B) these 1200 diver node sets were divided into 12 groups of 100 random driver sets with each group having a consistent number of high activity driver nodes ($N_{D_p}$) ranging between 1 and 12 across the 12 groups. Subfigures (C) and (D) each consist of 1200 diver node sets were divided into 12 groups of 100 random driver sets with each group having a consistent number of initially active driver nodes ($N_{D_a}$) ranging between 1 and 12 across the 12 groups.

\bibliography{reference}

\end{document}